\documentclass[superscriptaddress, pre, reprint]{revtex4-1}

\usepackage{amsmath}    
\usepackage{graphicx}   
\usepackage{graphics}   
\usepackage{verbatim}   
\usepackage{color}      
\usepackage{subfigure}  
\usepackage{hyperref}   
\usepackage{textcomp}   
\usepackage{float}
\usepackage[normalem]{ulem}

\begin{document}

\title{Energy partition at the collapse of spherical cavitation bubbles}

\author{M. Tinguely}
\affiliation{Laboratoire des Machines Hydrauliques, EPFL, 1007 Lausanne, Switzerland}
\author{D. Obreschkow}
\affiliation{Laboratoire des Machines Hydrauliques, EPFL, 1007 Lausanne, Switzerland}
\affiliation{ICRAR, The University of Western Australia, Crawley, WA 6009, Australia}
\author{P. Kobel}
\affiliation{Laboratoire des Machines Hydrauliques, EPFL, 1007 Lausanne, Switzerland}
\author{N. Dorsaz}
\affiliation{Department of Chemistry, University of Cambridge, Cambridge CB2 1EW, UK}
\author{A. de Bosset}
\affiliation{Laboratoire des Machines Hydrauliques, EPFL, 1007 Lausanne, Switzerland}
\author{M. Farhat}
\affiliation{Laboratoire des Machines Hydrauliques, EPFL, 1007 Lausanne, Switzerland}

\date{\today}

\begin{abstract}
Spherically collapsing cavitation bubbles produce a shock wave followed by a rebound bubble. Here we present a systematic investigation of the energy partition between the rebound and the shock. Highly spherical cavitation bubbles are produced in microgravity, which suppress the buoyant pressure gradient that otherwise deteriorates the sphericity of the bubbles. We measure the radius of the rebound bubble and estimate the shock energy as a function of the initial bubble radius (2-5.6 mm) and the liquid pressure (10-80 kPa). Those measurements uncover a systematic pressure dependence of the energy partition between rebound and shock. We demonstrate that these observations agree with a physical model relying on a first-order approximation of the liquid compressibility and an adiabatic treatment of the non-condensable gas inside the bubble. Using this model we find that the energy partition between rebound and shock is dictated by a single non-dimensional parameter $\xi = \Delta p\gamma^6/[{p_{g0}}^{1/\gamma} (\rho c^2)^{1-1/\gamma}]$, where $\Delta p=p_\infty-p_v$ is the driving pressure, $p_{\infty}$ is the static pressure in the liquid, $p_v$ is the vapor pressure, $p_{g0}$ is the pressure of the non-condensable gas at the maximal bubble radius, $\gamma$ is the adiabatic index of the non-condensable gas, $\rho$ is the liquid density, and $c$ is the speed of sound in the liquid.
\end{abstract}

\maketitle

\section{Introduction}
Research on cavitation is currently experiencing a rebirth within hydrodynamics. While traditionally associated with turbine erosion \cite{Rayleigh1917}, cavitation bubbles are now exploited as tools in surgery \cite{Vogel1997}, microchip cleaning \cite{Gale1995}, water treatment \cite{Kalumuck2000}, and microfluidics \cite{Yin2005,Dijkink2008}. This wide spectrum of new applications relies on the diversity of processes associated with the collapse of cavitation bubbles. Detailed studies revealed that these processes include (i) the formation of rebound bubbles \citep{Akhatov2001}, (ii) acoustic shocks \citep{Ohl1999}, (iii) thermal effects, leading to chemical reactions \citep{Suslick1990} and luminescence \citep{Barber1997,Brenner2002}, and (iv) micro-jets \citep{Blake1999,Obreschkow2006,Obreschkow2011b}. However, today there is no theory predicting the fraction of energy transferred into each of these processes. In the quest for such a theory, it seems wise to start with spherically collapsing bubbles, which produce no jets \cite{Obreschkow2011b}. We also note that thermal processes typically absorb negligible energy fractions \citep{Akhatov2001}. The problem then reduces to how the energy is split between rebound and shock in the spherical collapse.

This paper presents an experimental and theoretical investigation of the energy partition between rebound bubbles and shocks. We first describe the experiment, which uses a mirror-focused laser pulse in combination with micro-gravity conditions to produce bubbles of extremely high spherical symmetry. We then analyze measurements of the rebound sizes and the shock pressures of spherical bubbles produced in various experimental conditions. Interestingly, the energy ratio between rebound bubble and shock wave is found to decrease with the liquid pressure. We show that these observations can be explained using the Keller-Miksis collapse equation for a compressible liquid \cite{Keller1980} in combination with an adiabatic treatment of the non-condensable gas. Finally, we use this model to predict the energy partition between rebound and shock in a wide range of experimental conditions.

\section{Experimental setup}
\begin{figure}[h!]
\includegraphics[width=0.45\textwidth]{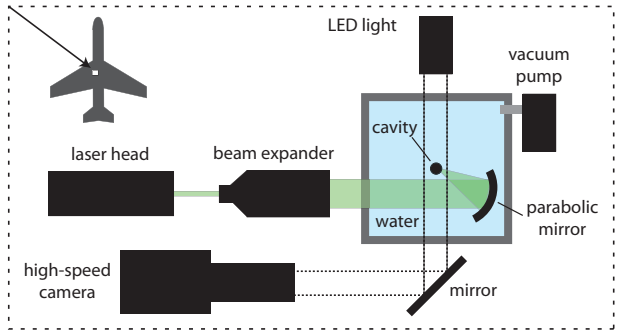}%
\caption{\label{SetupSketch} (Color online) Sketch of the experimental setup aboard the A300 zero-g aircraft.}
\end{figure}

The cavitation bubbles are obtained by focusing a high-energy laser in water (Fig. \ref{SetupSketch}). The laser source is a \emph{Q}-switched Nd:YAG laser (Quantel CFR 400) delivering pulses having 230 mJ maximum energy, 8 ns duration and a wavelength of 532 nm. The laser beam of 5 mm in diameter is expanded ten times before being focused by an off-axis parabolic mirror with a focal length of 54.5 mm, which is fixed inside the water container. The use of a parabolic mirror rather than an optical lens improves the focus by avoiding refraction and spherical aberration. The convergence angle is 53$^\circ$ and the focal point is located at the center of the water container. Owing to this large angle, the plasma generated at water breakdown is more compact and spherical than in previous studies \cite{Vogel1996}. The ensuing bubbles are so spherical that the faint hydrostatic pressure gradient due to gravity induces visible jets against the gravity vector \citep{Obreschkow2011b}. To avoid this source of asymmetry the experiment is performed in micro-gravity conditions [European Space Agency (ESA), 52nd parabolic flight campaign]. The flights consist of a total of 93 parabolic trajectories, flown by the aircraft A300 zero-g. Each parabola offers 20 s of microgravity (acceleration $<0.01 g$). Given those unique conditions, our experiment produces millimetric cavitation bubbles of extremely high sphericity.

The bubble is observed with a high-speed camera (Photron Fastcam SA1.1) at a rate of up to 250,000 frames/s with an exposure time of 370 ns. A 3W light-emitting diode (LED) light with a small opening angle of 6$^\circ$ is used to illuminate the bubble from the back and visualize the shock wave by shadowgraphy. The shock waves emitted at the generation and the collapse of the bubble are monitored by a piezo-resistive dynamic pressure sensor. The pressure in the vessel is controlled by a vacuum pump.

\section{Experimental results}
\begin{figure}[b]
\includegraphics[width=0.5\textwidth]{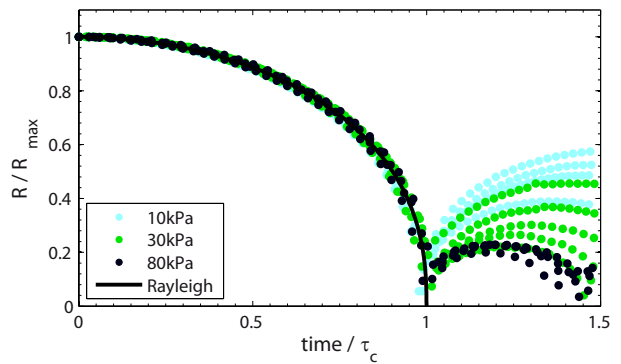}%
\caption{\label{Fig_RPstar} (Color online) The normalized radius for a representative selection of bubbles as a function of the normalized time, for different pressure levels $p_{\infty}$. The experimental data (dots) are consistent with the Rayleigh theory (solid black line).}
\end{figure}

\begin{figure*}[ht]
\includegraphics[width=1\textwidth]{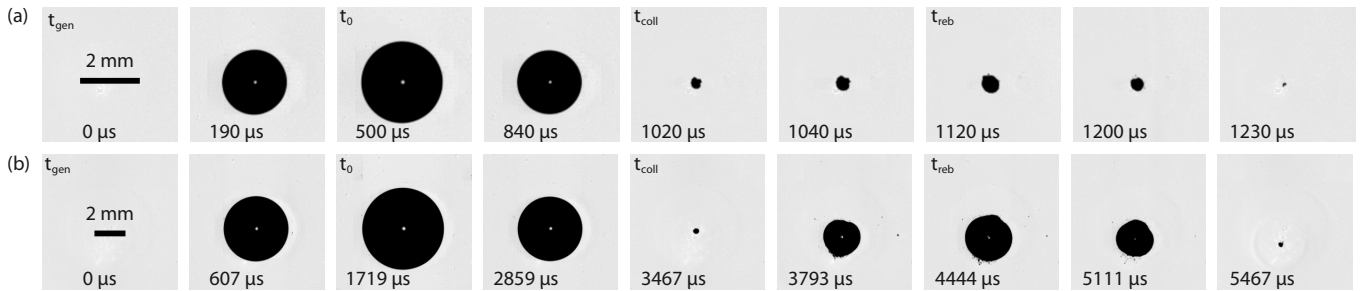}%
\caption{\label{Compare2B} Selected high-speed images of a cavitation bubble at two different water pressures. The images are scaled so that the bubble appears with the same normalized $R_{max}$ on the figure. (a) $p_\infty$ = 30 Pa, $R_{reb}/R_{max} = 0.22$ , (b) $p_\infty$ = 10 Pa, $R_{reb}/R_{max} = 0.57$.}
\end{figure*}

In the course of the flights, the bubble dynamics at three distinct water pressures $p_\infty$ (10, 30, and 80 kPa) is observed. For each pressure, the laser pulse energy is varied from 55 to 230 mJ, resulting in maximal bubble radii $R_{max}$ from 2 to 5.6 mm. Figure \ref{Fig_RPstar} shows the normalized radius $R/R_{max}$ for a representative selection of bubbles as a function of the normalized time $t/\tau_{c}$, where $\tau_{c}$ is the bubble collapse time according to Rayleigh theory \cite{Rayleigh1917}, $\tau_{c} = 0.915 R_{max} \sqrt{\rho/\Delta p}$ with $\rho$ being the density of the liquid and $\Delta p$ being the ``driving pressure,'' i.e., the difference between the static liquid pressure $p_\infty$ and the pressure $p_v$ of the condensable vapor inside the bubble.
The value of $p_v$ is calculated with the Antoine equation from the temperature of the water measured for each of the three flight days. The three temperatures are, respectively, 16.8, 23.9, and 20.9 $^\circ$C, corresponding to $p_v$ of 1910, 2950, and 2460 Pa.
All the curves are remarkably superposed during the first collapse, and are consistent with the Rayleigh theory (solid line in the figure).
However, the dynamics of the rebound is very different depending on the pressure in the liquid $p_{\infty}$ (see also Fig. \ref{Compare2B}). The high-speed movies reveal that the normalized maximum radius of the first rebound bubble $R_{reb}/R_{max}$ decreases with $p_\infty$. To interpret this result in terms of energy, we calculate the potential energy of a bubble as \citep{Obreschkow2006}
\begin{equation} \label{eq_Epot_liq} \displaystyle
	E_{pot} = \int_0^R{4\pi r^2\Delta p\,{\rm d}r} = \frac{4\pi}{3}R^3 \Delta p,
\end{equation}
where $R$ is the bubble radius. In particular, we define the initial bubble energy $E_0$ and the rebound energy $E_{reb}$ as
\begin{equation}\label{eq_Epot2}
\begin{array}{lcr} \displaystyle
	E_{0}\!=\!\frac{4\pi}{3}R_{max}^3 \, \Delta p &\rm{and} & \displaystyle E_{reb}\! =\! \frac{4 \pi}{3}\! R_{reb}^3 \, \Delta p \,.
\end{array}
\end{equation}

\begin{figure}[b]
\includegraphics[width=0.5\textwidth]{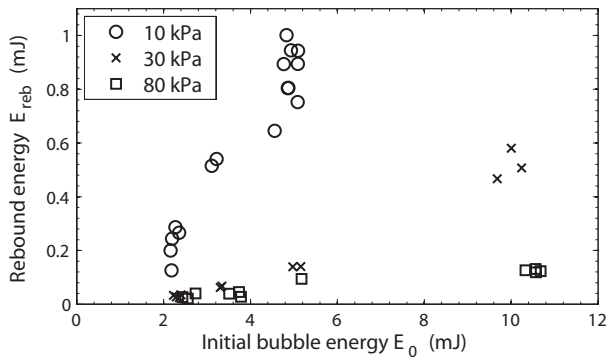}
\caption{\label{Ereb_Emax}Measured potential energy of the rebound bubble as a function of the initial bubble energy, for different pressure levels $p_\infty$.}
\end{figure}

Figure \ref{Ereb_Emax} reveals that $E_{reb}$ scales with $E_{0}$ for given liquid pressures $p_\infty$.

To complete the picture, we consider the energy carried away by the spherical shock produced at the first bubble collapse. Given a shock pressure $p(t)$, measured at a distance $d$ from the bubble center, the shock energy is given by \cite{Vogel1996}
\begin{equation}\label{eq_SW}
	E_{SW} = \frac{4 \pi d^2}{\rho c} \int{p(t)^2 {\rm d}t},
\end{equation}
where $\rho$ is the water density and $c$ is the speed of sound in water. In our experiment, the duration of the shock transition, i.e.,~the characteristic time scale of $p(t)$ ($<$100 ns), is much shorter than the characteristic response time (10 $\mu s$) of the pressure sensor. Nevertheless, a rough estimation of the shock energy remains possible under the assumption of a linear response. Explicitly, if we define $h(t)$ as the sensor's impulse response, the response of the sensor $s(t)$ is expressed as $s(t) = h(t)\ast p(t)$, where ``$\ast$'' denotes the convolution. We assume that the pressure $p(t)$ has a universal shape in the sense that $p(t)=p_{max} \, \tilde{p}(t)$, where $\tilde{p}(t)$ is the same function for all bubbles. The signal can then be expressed as $s(t) = h(t)\ast p_{max}\,\tilde{p}(t)$, and hence $\int s(t) {\rm d}t = p_{max}\int h(t)\ast\tilde{p}(t) {\rm d}t\propto p_{max}$. In other words, $p_{max}$ is proportional to the integrated response.
Substituting into Eq.~(\ref{eq_SW}), we finally obtain \footnote{See Supplemental Material at [...] for details on the validation of the assumptions.}
\begin{equation}\label{eq_SW2}
	E_{SW} \propto \int{p_{max}^2 \,\tilde{p}(t)^2{\rm d}t} \propto p_{max}^2 \propto \left(\int s(t){\rm d}t\right)^2.
\end{equation}

\begin{figure}[b]
\includegraphics[width=0.5\textwidth]{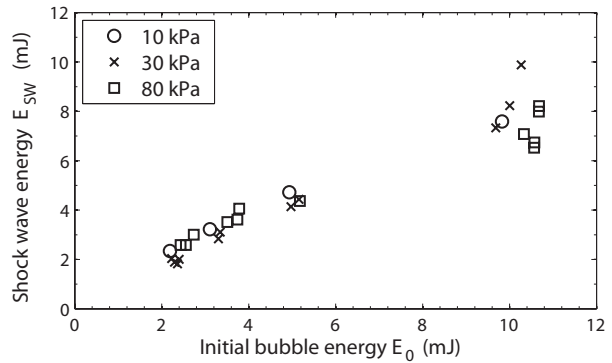}
\caption{\label{ESW_Emax}Estimated energy in the shock wave as a function of the initial bubble energy, for different pressure levels $p_\infty$.\\
$\!$}
\end{figure}

The constant of proportionality in Eq.~(\ref{eq_SW2}), which is unknown, is estimated such that the shock energy $E_{SW}$ equals the initial energy $E_0$ in the extreme cases, where only a negligible rebound bubble is observed. The measured shock energy $E_{SW}$ versus the initial potential energy $E_{0}$ are presented in Fig.~\ref{ESW_Emax}. Unlike the rebound energy $E_{reb}$ (Fig.~\ref{Ereb_Emax}), $E_{SW}$ is roughly proportional to $E_{0}$ with no clear dependence on pressure. As we will show below, this is well explained by the fact that our experiments all lie in a ``shock-saturated'' regime, where the shock absorbs most of the available energy ($E_{SW}\approx E_0$).

\section{Theoretical model}
Hereafter, a theoretical model is developed to compute the energies of the rebound bubble and the shock wave as a function of various experimental parameters. The standard model for the evolution of spherical cavitation bubbles, i.e.,~the Rayleigh-Plesset equation, cannot produce rebound bubbles and shock waves. To calculate the rebound motion it is necessary to include a non-condensable gas inside the bubble. We here assume that this gas is compressed and decompressed adiabatically, i.e., that is without heat transfer across the bubble surface.
According to the adiabatic theory, the pressure $p_g(t)$ of this non-condensable gas is then given by\cite{Brennen1995}
\begin{equation}\label{eq_pg}
 p_g = p_{g0}\left(\frac{R_{max}}{R}\right)^{3\gamma},
\end{equation}
where $p_{g0}$ is the pressure at the maximal initial bubble radius $R_{max}$, $R(t)$ is the evolving bubble radius, and $\gamma$ is the adiabatic index also known as ``heat capacity ratio.'' Second, to incorporate shock waves, we require a model for the bubble evolution in a compressible liquid. We here use the Keller-Miksis model \cite{Keller1980}, which is an extension of the Rayleigh equation to compressible liquids, accurate to first order in the speed of sound $c$. As shown by Prosperetti \cite{Prosperetti1987} this model belongs to a more general class of first-order models and can be rewritten as
\begin{equation}\label{eq_prospe}
\ddot{R} = \frac{(p_g-\Delta p)(1+\tilde{v})+R\dot{p}_g/c-(3-\tilde{v})\dot{R}^2\rho/2}{(1-\tilde{v})R \rho},
\end{equation}
where $\tilde{v}(t)\equiv \dot{R}(t)/c$. Note that we deliberately neglect the effects of surface tension and viscosity for two reasons. First, these effects are quite irrelevant for the large bubbles in our experiment. Second, surface tension and viscosity are generally insignificant at the last stage of the bubble collapse, since inertial forces increase more rapidly than viscous forces and surface tension as $R(t)\rightarrow 0$. The latter can therefore be neglected to calculate rebounds and shocks.

Equations.~(\ref{eq_pg}) and (\ref{eq_prospe}), fitted with the initial conditions $R(0)=R_{max}$, $\dot{R}(0)=0$, $p_g(0)=p_{g0}$, and $\dot{p}_g(0)=0$, constitute a model for the collapse and the rebound of a spherical bubble, while including compression waves (shocks). We use the Runge-Kutta method to solve this model numerically. The radius $R(t)$ is calculated as the bubble first collapses and then rebounds until it reaches its maximal rebound radius $R_{reb}$.

Given a time solution of Eqs.~(\ref{eq_pg}) and (\ref{eq_prospe}) we can then calculate various energies. The initial bubble energy $E_0$ and the energy of the rebound bubble $E_{reb}$ are computed directly using Eq.~(\ref{eq_Epot2}). It is important to note that the temperature of the non-condensable gas changes during the adiabatic compression and decompression. The gas temperature at the rebound point is different from the initial temperature. Hence the internal energy $U=(4\pi/3)R^3p_g/(\gamma-1)$ of the non-condensable gas changes. We can calculate this energy change $\Delta U$ simply by subtracting the final value of $U$ from the initial one,
\begin{equation}\label{eq_DeltaU1}
	\Delta U = \frac{4\pi}{3(\gamma-1)}\left(p_{g0}R_{max}^3-p_{g,reb}R_{reb}^3\right).
\end{equation}
The adiabatic nature of the process implies that $\Delta U$ must be equal to the total work done by the liquid onto the non-condensable gas. This work can be calculated as
\begin{equation}\label{eq_DeltaU2}
	\Delta U = \int\delta W = -\int p_g{\rm d}V = -\int 4\pi R^2\dot{R}\,p_g\,{\rm d}t,
\end{equation}
where the time integral runs from the initial bubble radius through the collapse point to the maximal rebound radius. To check the accuracy of our numerical solution we compute $\Delta U$ using both Eqs.~(\ref{eq_DeltaU1}) and (\ref{eq_DeltaU2}).

Given $\Delta U$, the initial energy $E_0$, and the potential energy of the rebound $E_{reb}$, the compression energy of the shock wave $E_{SW}$ can be computed from energy conservation as
\begin{equation}\label{eq_ESW}
	E_{SW} = E_0-E_{reb}-\Delta U.
\end{equation}
Finally, we introduce the energy fractions
\begin{equation}
	\epsilon_{reb}\equiv E_{reb}/E_0,~~\epsilon_{SW}\equiv E_{SW}/E_0,~~\epsilon_U\equiv\Delta U/E_0.
\end{equation}
Equation~(\ref{eq_ESW}) implies the normalization $\epsilon_{reb}+\epsilon_{SW}+\epsilon_U=1$.

\begin{figure}[b]
\includegraphics[width = 0.5 \textwidth]{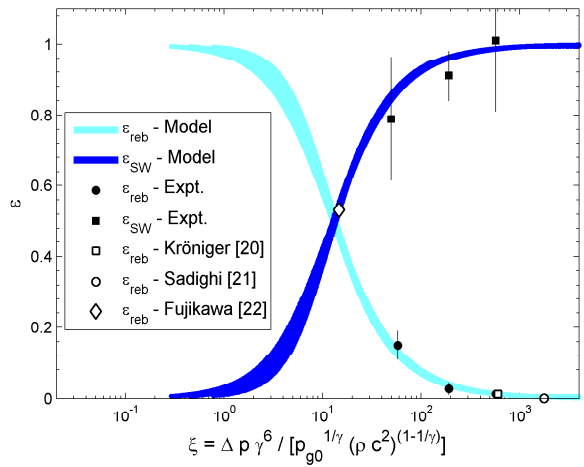}
\caption{\label{ThMo_Xi} (Color online) Fraction of energy in the rebound $\epsilon_{reb}$ and in the shock wave $\epsilon_{SW}$ as a function of the non-dimensional parameter $\xi$. The solid curves are the results from the theoretical model. The discrete black symbols are the values obtained experimentally, along with the measurement error bars. The white symbols are data extracted from the literature.}
\end{figure}

How do $\epsilon_{reb}$, $\epsilon_{SW}$, and $\epsilon_U$ depend on the six model parameters $R_{max}$, $\Delta p$, $p_{g0}$, $\gamma$, $\rho$, and $c$ ? We first note that the four energies $E_0$, $E_{reb}$, $E_{SW}$, and $\Delta U$ all scale as $R_{max}^3$. This can be shown by rewriting the model as a function of the normalized radius $r(t)\equiv R(t)/R_{max}$. Therefore $\epsilon_{reb}$, $\epsilon_{SW}$, and $\epsilon_U$ are independent of $R_{max}$. To test the remaining five model parameters we ran $2.7\times10^5$ independent computations of $\epsilon_{reb}$, $\epsilon_{SW}$, and $\epsilon_U$ by taking logarithmically spaced parameters from the following intervals: $\Delta p\in\rm[1,100]~kPa$, $p_{g0}\in\rm[0.1,100]~Pa$, $\rho\in\rm[500,1500]~kg\,m^{-3}$, $c\in\rm[1000,2000]~m\,s^{-1}$, $\gamma\in[1.3,1.5]$. By systematically studying the variation of $\epsilon_{reb}$, $\epsilon_{SW}$, and $\epsilon_U$ as a function of the five parameters, we can draw two main conclusions. First, the internal energy fraction is negligible because $\epsilon_U<0.01$ in all situations. Second, all variations of $\epsilon_{reb}$ and $\epsilon_{SW}$ as a function of the five model parameters $\Delta p$, $p_{g0}$, $\gamma$, $\rho$, and $c$ can be explained by using a single non-dimensional parameter
\begin{equation}\label{eq_xi}
	\xi = \frac{\Delta p\gamma^6}{{p_{g0}}^{1/\gamma} (\rho c^2)^{1-1/\gamma}}.
\end{equation}
In fact, Fig.~(\ref{ThMo_Xi}) shows the $2.7\cdot10^5$ values of $\epsilon_{reb}$ and $\epsilon_{SW}$ as a function of $\xi$, revealing a tight correlation. The parameter $\xi$ was found by first constructing the non-dimensional parameter $\Delta p\,(p_{g0})^a (\rho c^2)^{-a-1}$ from the four dimensional parameters $\Delta p$, $p_{g0}$, $\rho$, and $c$. The computed results are then grouped depending on the value of $\gamma$. $a$ is determined for each group as the value that maximizes the Pearson correlation coefficient for $\epsilon_{reb} \in [0.2, 0.8]$. We restrict $\epsilon_{reb}$ to the interval where a small variation in $\xi$ leads to a large variation of $\epsilon_{reb}$, thus where we want the relation to be the most univocal. The values of $a$ obtained depend on $\gamma$ as $a=1/\gamma$ with an error of $\pm 10\%$. As the curves obtained for each value of $\gamma$ are still horizontally shifted, the second step is to introduce the factor $\gamma^\beta$. $\beta=6$ is then determined by maximizing the Pearson correlation coefficient on $\epsilon_{reb} \in [0.2, 0.8]$ for all values of $\gamma$.

\section{Discussion}

\subsection{Comparison between model predictions and experiment} \label{sec_comparison_model_exp}
The theoretical model allows us to explain why, according to our experimental results, the energy of the rebound depends on the pressure in the liquid while the energy of the shock wave seems to scale with the initial potential energy only. However, to compare the experimental results with the theoretical ones, we need a value for $p_{g0}$ in addition to the measured $\Delta p$ and $R_{max}$ and the known $\rho$, $c$, and $\gamma$. Since $p_{g0}$ is not directly measurable, we simply assume this pressure to be constant. We estimated its value by fitting the model to the experimental results. For each measurement, the value of $p_{g0}$ leading to the observed $R_{reb}$ is calculated with an iterative process. The results are averaged and we obtain $\overline{p_{g0}}= \rm{7.0} \pm \rm{3.5}$Pa. The relatively small variance a posteriori justifies the assumption of a constant value for $p_{g0}$. The experimental points are plotted in Fig. \ref{ThMo_Xi}, where the values of $\xi$ are calculated using $\overline{p_{g0}}$. We observe that all our experimental data lies in a regime where $\epsilon_{SW}\approx1$. So when $\xi$ varies because of the change of $\Delta p$, the relative difference is important for $\epsilon_{reb}$ but not for $\epsilon_{SW}$. As $\epsilon_{reb}$ and $\epsilon_{SW}$ represent the slopes of the curves in Figs. \ref{Ereb_Emax} and \ref{ESW_Emax}, respectively, the difference in $p_{\infty}$ is significant for the rebound, but insignificant for the shock.

We observe, both theoretically and experimentally, the relation $E_{reb} + E_{SW} = E_{0}$. The results using the theoretical model show that $\Delta U$ in Eq.(\ref{eq_ESW}) is negligible, which implies $E_{reb} + E_{SW} = E_{0}$. And when adding the experimental data from Figs. \ref{Ereb_Emax} and \ref{ESW_Emax}, we obtain $E_{reb} + E_{SW} = E_0$, within a scatter of $\pm 20\% $. This scatter is reasonable considering the uncertainty introduced with the rough estimation of $E_{SW}$.

\subsection{Comparison with earlier work}
The main issue with the treatment of the rebound is that the pressure of non-condensable gas $p_{g0}$ is usually needed in the equation of motion. As $p_{g0}$ is not measurable and its origin is not clear yet, it is difficult to estimate it and thus to validate a model. So in a concern of evaluating our theoretical model, we look at previous studies for comparison. In the experimental and numerical work of Kr\"{o}ninger \emph{et al.} \cite{Kroninger2009}, and in the numerical work of Sadighi-Bonabi \emph{et al.} \cite{Sadighi2012} and Fujikawa \emph{et al.} \cite{Fujikawa1980}, we found estimates of $p_{g0}$ or enough information to obtain them. The data extracted from these articles are plotted in Fig. \ref{ThMo_Xi}. We observe that despite the different treatments of the thermodynamics inside the bubble (Ref. \cite{Kroninger2009} used a van der Waals equation, Ref. \cite{Sadighi2012} considered hydrochemical reactions, and Ref. \cite{Fujikawa1980} considered conductive heat transfer and condensation or evaporation) our model reproduces reasonably well their results. Yet, the drawback of our model is that the temperature at the collapse is overestimated because of the neglected thermal transport. This could be improved by the addition of heat transfer or chemical reactions, but at the cost of the simplicity of the model.

Akhatov \emph{et al.} \cite{Akhatov2001} propose a mathematical model supported by experimental measurements of the rebound of a spherical cavitation bubble. Because of the difference in the model used (the pressure of non-condensable gas is derived from a phase transition equation), we could not derive a value for $p_{g0}$ for quantitative comparison of our results. Nonetheless, qualitatively, the conclusions are the same. Akhatov \emph{et al.} observed that the ratio between the radius of the rebound and the initial bubble is constant when only varying the initial radius of the bubble, which confirms the univocal relation between $\xi$ and $\epsilon_{reb}$. They also showed numerically that for given experimental conditions, when the concentration of the non-condensable gas in the bubble increases, the radius of the rebound bubble increases too. This also agrees with our conclusions. Indeed, the increase of the concentration of non-condensable gas means a smaller value of $\xi$ which implies, according to Fig. \ref{ThMo_Xi}, an increase of $\epsilon_{reb}$ and thus of the rebound radius.

\subsection{Negligible role of gravity}
\begin{figure}[h]
\includegraphics[width = 0.5 \textwidth]{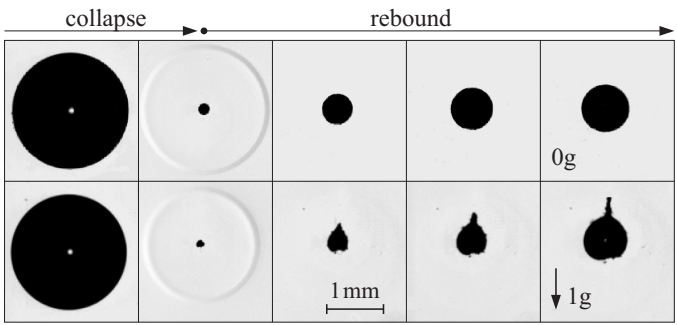}
\caption{\label{Fig_bubble_jet} Collapse and rebound of a bubble in 0$g$ (upper) and 1$g$ (lower). Note the shock visible at the collapse, and the vapor jet on the rebound for 1$g$ \cite{Obreschkow2011b}. }
\end{figure}

We have already demonstrated (see Obreschkow \emph{et al.} \citep{Obreschkow2011b}) that gravity can affect the collapse of a cavitation bubble in the form of the occurrence of a vapor jet (see Fig. \ref{Fig_bubble_jet}). The volume of the vapor jet normalized to the maximum volume of the rebound was found to be proportional to the non-dimensional parameter $\zeta = |\nabla p| R_0 / \Delta p$, where $\nabla p$ is the hydrostatic pressure gradient.
Therefore we also performed the experiments in this paper with the same parameters at normal gravity ($1 g$) and hypergravity ($1.8 g$). The values of the non-dimensional parameter were $\zeta \in [\rm{2.5 \cdot 10^{-3}}, \rm{7 \cdot 10^{-3}}]$. Unlike for the vapor jet, we do not observe a significant difference in the energy partition when the gravity changes. The relative difference between the values of $\epsilon_{reb}$ at 0$g$ and at $\geq$1$g$ are smaller than the standard deviations of the measurements at 0$g$.
 We deduce that the energy transferred into the vapor jet is negligible compared to the energy in the rebound and in the shock. In consequence, the results of this paper also apply to bubbles collapsing in a hydrostatic pressure gradient for $\zeta < \rm{7 \cdot 10^{-3}}$.

Note that in the studies cited here \citep{Akhatov2001, Fujikawa1980, Kroninger2009, Sadighi2012} and plotted in Fig.\ref{ThMo_Xi}, the non-dimensional parameter is $\zeta \ll \rm{7 \cdot 10^{-3}}$: We found respectively $\rm{1.5 \cdot 10^{-4}}$, $\rm{7.5 \cdot 10^{-5}}$, $\rm{3.0 \cdot 10^{-6}}$, and $\rm{1.4 \cdot 10^{-4}}$. We thus consider that the comparison of our results with these data is justified.

\subsection{Implications}
A systematic experimental and theoretical investigation of the rebound and shock energy at the collapse of a spherical cavitation bubble is presented.  This led us to identify a single non-dimensional parameter $\xi$, which links the experimental conditions to the fraction of energy in the rebound bubble and in the shock wave. This finding has important implications for many engineering applications. Depending on the desired requirements, we can tune the experimental parameters in order to obtain a value for $\xi$ that favors either the rebound or the shock. For example in micro-pumping, an enhanced rebound is desired to increase efficiency, because the volume of the bubble affects the displacement of the liquid \cite{Dijkink2008}. Conversely, for applications that rely on cavitation erosion, stronger shocks would be preferred to accelerate the process.

The methodology presented here to estimate $p_{g0}$ also opens perspectives for understanding the origin of the non-condensable gas in the bubble at the collapse. So far the non-condensable gas has been assumed to be a combination of trapped vapor, laser breakdown products, and gas initially present in the water \cite{Akhatov2001}. A method to verify this would be to systematically vary the experimental conditions and assess their effect on $p_{g0}$. Our model could then be used to extract the values of $p_{g0}$ by fitting the experimental results with the theoretical ones. In the same line of thought, we could estimate water properties, such as the concentration of dissolved gas and nuclei, solely based on observing the rebound of natural or artificially generated cavitation bubbles. These results could be complemented with observations of extreme cases such as cryogenic fluids, where the experimental parameters (``driving pressures'' $\Delta p$, adiabatic index $\gamma$, density $\rho$, and speed of sound $c$) are very different compared to the case of a bubble in water \citep{Roche1996,Tomita2000}.

\section{conclusion}
We observe experimentally that the pressure of the liquid affects the energy partition between the rebound and the shock wave at the collapse of a highly spherical laser-induced bubble. This unique experiment is performed in microgravity conditions in order to avoid the formation of a microjet due to the hydrostatic pressure gradient at the collapse of the millimetric bubble. A theoretical model for the collapse of spherical bubbles is proposed, relying on a compressible equation of motion and an adiabatic treatment of the non-condensable gas inside the bubble. The partition of the energy between rebound and shock is calculated for a wide range of parameters. It appears that, in addition to the pressure in the liquid, the physical properties of the liquid and the pressure of non-condensable gas inside the bubble affect the energy partition. These parameters can be combined into a single non-dimensional parameter $\xi$, which dictates the energy partition.

The ability to predict the energy partition between rebound and shock is valuable in many engineering applications. The operating conditions can be adjusted to favor rebound or shock depending on the requirements. Using the method developed in this paper, it is also possible to estimate the pressure of the non-condensable gas in the bubble by fitting, with an iterative process, the experimental observations of the radius of the rebound bubble with the theoretical results. However, as the pressure of the non-condensable gas is not directly measurable, this method still has to be quantitatively validated with further experiments.

\section*{Acknowledgements}
The authors wish to thank the Swiss National Science Foundation for their support under Grants No. 200020-116641 and No. PBELP2-130895, and the European Space Agency (ESA) for the access to the parabolic flight campaign (52nd ESA PFC).\\


%

\end{document}